\documentclass[aps,pre,onecolumn,showpacs]{revtex4}
\usepackage{amsmath}
\usepackage{epsfig}
\usepackage{amssymb}
\usepackage{dcolumn}
\usepackage{graphicx}
\usepackage{dcolumn}

\usepackage{amssymb}
\newcommand{\sn}{\mathop{\mbox{sn}}}
\newcommand{\dn}{\mathop{\mbox{dn}}}

\begin{document}
\date{\today}

\title{Lie Symmetries, qualitative analysis and exact solutions of nonlinear Schr\"odinger equations
with inhomogeneous nonlinearities}

\author{Juan Belmonte-Beitia$^{1,3}$}
\email{juan.belmonte@uclm.es}

\author{V. M. P\'erez-Garc\'{\i}a$^{1,3}$}
\email{victor.perezgarcia@uclm.es}

\author{V. Vekslerchik$^{1,3}$}
\email{vadym@vekslerchik.uclm.es}
\homepage{http://matematicas.uclm.es/nlwaves}

\author{P. J. Torres$^{2}$}
\email{ptorres@ugr.es}

\affiliation{
$^1$ Departamento de Matem\'aticas, E. T. S. de Ingenieros Industriales, 
Universidad de Castilla-La Mancha 13071 Ciudad Real, Spain.
\\
$^2$Departamento
de Matem\'atica Aplicada. Facultad de Ciencias. Universidad de Granada
Campus de Fuentenueva s/n, 18071 Granada, Spain.
\\
$^3$Instituto de Matem\'atica Aplicada a la Ciencia y la Ingenier\'{\i}a (IMACI),
Universidad de Castilla-La Mancha, 13071 Ciudad Real,
Spain
}

\begin{abstract}
 Using Lie group theory and canonical transformations, we construct explicit solutions of  nonlinear Schr\"odinger equations with  spatially inhomogeneous nonlinearities. We present the general theory, use it to  study different examples and use the qualitative theory of dynamical systems to obtain some properties of these solutions.
 \end{abstract}


\maketitle

\section{Introduction}\label{s1}

The Nonlinear Schr\"{o}dinger Equation (NLSE) in its many versions is one of the most important models of mathematical physics, with applications to different fields \cite{Vazquez} as for example in
semiconductor electronics \cite{Soler,Soler2}, nonlinear optics
 \cite{Kivshar}, photonics \cite{Hasegawa}, plasma physics
\cite{Dodd}, fundamentation of quantum mechanics
\cite{fundamentals}, dynamics of accelerators \cite{Fedele},
mean-field theory of Bose-Einstein condensates \cite{Dalfovo,VV} or
  biomolecule dynamics \cite{Davidov} to cite only a few examples. In some of these fields and many others, the NLSE
appears as an asymptotic limit for a slowly varying dispersive
wave envelope propagating in a nonlinear medium \cite{scott}. Moreover, the range of applicability is large because of the well-known universality of this equation \cite{Bambusi}.

The study of these equations has served as a catalyzer of the
development of new ideas or even mathematical concepts such as solitons \cite{Zakharov}
or singularities in partial differential equations \cite{Sulembook,SIAMFibich}.

 In the last years there has been an increased interest in a one-dimensional nonlinear Schr\"odinger equation with inhomogeneous nonlinearity (INLSE):
 \begin{equation}
 \label{NLSEinh}
 i\psi_{t} = -\psi_{xx}+V(x)\psi + g(x)\left\lvert\psi\right\rvert^{2}\psi,
 \end{equation}
 with $x\in \mathbb{R}$,  $V(x)$ is an external potential and $g(x)$ describes the spatial modulation of the nonlinearity. This equation arises in different physical contexts such as nonlinear optics and  dynamics of Bose-Einstein condensates. But it is in the later field where the possibility of using the Feschbach resonance management techniques to modify spatially the collisional interactions between atoms
 \cite{Victor1,Victor2,Garnier,Panos,Garnier2,Primatarowa,Panos2,Hu}
the one which has motivated a lot of theoretical research in the last few years focusing on  questions of direct applicability to experiments. Different aspects of the dynamics of solitons in these contexts have been studied such as the emission of solitons \cite{Victor1,Victor2} and the propagation of solitons when the space modulation of the nonlinearity is a random \cite{Garnier}, periodic \cite{Boris,Panos2}, linear \cite{Panos} or localized function
\cite{Primatarowa}.
Although many exact solutions of the NLSE with spatially homogeneous nonlinearities and without potentials ($V=0$) have been known for a long time,  the problem of finding exact solutions even of the NLS with homogeneous nonlinearities and general potentials is a very difficult one.

 In this paper, using Lie symmetries we find general classes of potentials $V(x)$ and nonlinearity functions $g(x)$
 for which exact solutions can be constructed by combining solutions of the integrable NLS and solvable potentials $V(x)$. The basic idea of the Lie symmetries method is to study the invariance properties of given differential equations under continuous groups of transformations.
 This method  has been applied successfully to different equations, such as, for example, differential equations that model anharmonic oscillators \cite{Leach,Leach1} and Madelung fluid equations \cite{Baumann}. In Ref. \cite{Nuestro} we have presented some examples of this methodology of specific physical interest. Here we complement that analysis by presenting  the general theory, provide more examples, study the case of asymmetric solutions and use qualitative theory of dynamical systems to provide a much more complete analysis of the method and its applications to equations of physical relevance.

 The paper is organized as follows. In Section 2, we introduce the general theory of the Lie symmetry analysis for ordinary differential equations and particularize it for our model problem: the nonlinear Schr\"odinger equation with an inhomogeneous nonlinearity. In Section 3, we study the canonical transformations of the INLSE. In Section 4, we present the connection between the NLSE and the INLSE. In Section 5, we use the method to construct explicit solutions of the stationary  nonlinear Schr\"odinger equation with an inhomogeneous nonlinearity and  study the qualitative behaviour of the NLSE and its qualitative connection with the INLSE in different examples. Finally, in Section 6, we present asymmetric solutions of the INLSE.  To our knowledge, this is the first time that such solutions are calculated.

\section{General Theory of Lie symmetries}\label{s2}

In this paper we will look for localized stationary solutions of Eq. (\ref{NLSEinh}), which are of the form
\begin{equation}
\psi(x,t)=u(x) e^{-i\lambda t}, \nonumber
\end{equation}
which satisfy the following nonlinear eigenvalue problem
 \begin{subequations}
  \begin{eqnarray}
 \label{estacionario}
  -u_{xx}+V(x)u+g(x)u^{3}&=&\lambda u, \label{cuca}\\
   \lim_{|x|\rightarrow\infty}u(x)&=&0.
 \end{eqnarray}
 \end{subequations}
By definition \cite{Bluman,Olver}, a second-order differential equation $A(x,u,u',u'')=0 $
possesses a Lie group of point transformations or Lie point symmetry of the form
\begin{equation}
M=\xi(x,u) \partial/\partial x+\eta(x,u)\partial/\partial u, \nonumber
\end{equation}
 if the action of the second extension of $M$, i.e. $M^{(2)}$ on $A$ is equal to zero, i.e.
\begin{multline}
M^{(2)} A(x,u,u',u'') = \left[\xi(x,u)\frac{\partial}{\partial x}+\eta(x,u)\frac{\partial}{\partial u}+ \right. \\ \left. \eta^{(1)}(x,u)\frac{\partial}{\partial u'}+
\eta^{(2)}(x,u)\frac{\partial}{\partial u''}\right] A(x,u,u',u'') = 0, \nonumber
\end{multline}
with $\eta^{(k)}$ given by
\begin{equation}\label{coeficienteds}
\eta^{(k)}(x,u,u',u'',...,u^{k})=\frac{D\eta^{(k-1)}}{Dx}-u^{(k)}\frac{D\xi(x,u)}{Dx}, \ \ \ k=1,2,... \nonumber
\end{equation}
where $\eta^{(0)}=\eta(x,u)$ and $D/Dx$ is the total derivative, i.e.
\begin{equation}
\frac{D}{Dx}=\frac{\partial}{\partial x}+u'\frac{\partial}{\partial u}+u''\frac{\partial}{\partial u'}+....+u^{(n+1)}\frac{\partial}{\partial u^{(n)}}+... \nonumber
\end{equation}
For example, $\eta^{(1)}$ is equal to
\begin{equation}
\eta^{(1)}=\eta_{x}+[\eta_{u}-\xi_{x}]u_{x}-\xi_{u}(u_{x})^{2}. \nonumber
\end{equation}
In our case, $A(x,u,u_{x},u_{xx})$ is given by
\begin{equation}
A(x,u,u_{x},u_{xx})=-u_{xx}+f(x,u), \nonumber
\end{equation}
where $f(x,u)=V(x)u+g(x)u^{3}-\lambda u$, and the action of the operator $M^{(2)}$ on $A(x,u,u_{x},u_{xx})$ leads to a polynomial equation in $u_{x}$. By equating coefficients of powers of $u_{x}$, one obtains
\begin{subequations}
\begin{eqnarray}
\label{relaciones1}
\xi_{uu}&=&0,
\\ \label{relaciones2}
\eta_{uu}-2\xi_{u x}&=&0,
\\ \label{relaciones3}
2\eta_{xu}-\xi_{xx}-3f\xi_{u}&=&0,
\\ \label{relaciones4}
\eta_{xx}-\xi f_{x}-\eta f_{u}+\eta_{u}f-2\xi_{x}f&=&0.
\end{eqnarray}
\end{subequations}
Integrating Eqs. (\ref{relaciones1}) and (\ref{relaciones2}), we get
\begin{equation}\label{ec1}
\xi(x,u)=a(x)u+b(x), \quad \quad \eta(x,u)=a'(x)u^{2}(x)+c(x)u+d(x).
\end{equation}
Substituting these expressions into Eq. (\ref{relaciones3}) we obtain
\begin{equation}\label{ec2}
2c'(x)=b''(x), \quad \quad a(x)=0.
\end{equation}
Finally, substituting Eqs. (\ref{ec1}) and (\ref{ec2}) in  Eq. (\ref{relaciones4}), we get
\begin{subequations}
\begin{eqnarray}\label{r1}
\xi(x,u)&=&b(x),
\\ \label{r2}
\eta(x,u)&=&c(x)u,
\\ \label{r3}
c''(x)-b(x)V'(x)-2b'(x) \left(V(x)-\lambda\right)&=&0,
\\ \label{r4}
2c(x)g(x)+b(x)g'(x)+2b'(x)g(x)&=&0.
\end{eqnarray}
\end{subequations}
Also the substitution of Eq. (\ref{ec2}) in Eq. (\ref{r4}) gives
\begin{equation}
g(x)  =  g_{0}b^{-3}(x)e^{-2C\int_{0}^{x}1/b(s)ds}. \nonumber
\end{equation}
where $g_{0}$ and $C$ are arbitrary constants.
Summarizing the previous calculations, the Lie point symmetry is of the form
\begin{equation}
\label{simetria}
M=b(x)\frac{\partial}{\partial x}+c(x)u\frac{\partial}{\partial u},
\end{equation}
where
\begin{subequations}\label{relaciones}
\begin{eqnarray}\label{relacionesa}
g(x) & = & g_{0}b^{-3}(x)e^{-2C\int_{0}^{x}1/b(s)ds},\\ \label{relacionesb}
c(x) & = & \tfrac{1}{2}b'(x)+C,\\  \label{relacionesc}
c''(x)-b(x)V'(x)-2b'(x) \left(V(x)-\lambda\right)&=&0. 
\end{eqnarray}
\end{subequations}
 Eqs. (\ref{relaciones}) allow us to construct pairs $\{ V(x), g(x) \}$ for which a Lie point symmetry exists. Thus given either $g(x)$ or $V(x)$, in principle we can choose the other in order to satisfy Eqs. (\ref{relaciones}). In what follows, we will study the implications of the existence of this Lie symmetry.

\section{Canonical transformations and invariants.}\label{s3}

It is known \cite{Leach}, that the invariance of the energy is associated to the translational invariance.  The generator of such a transformation is of the form $M=\partial/\partial X$. To use this fact, we define the transformation from variables $(x,u)$ to new variables $(X,U)$
\begin{equation}
\label{transformaciones}
X=h(x),\qquad U=n(x)u,
\end{equation}
where $h(x)$ and $n(x)$ will be determined by requiring
 that a conservation law of energy type $M=\partial/\partial X$ exists
in the canonical variables. In fact, using Eq. \eqref{transformaciones}, we get
\begin{eqnarray}\label{ex1}
\frac{\partial}{\partial u}=n(x)\frac{\partial}{\partial U}, \\ \label{ex2}
\frac{\partial}{\partial x}=n'(x)u\frac{\partial}{\partial U}+h'(x)\frac{\partial}{\partial X}.
\end{eqnarray}
Inserting the expressions (\ref{ex1}) and (\ref{ex2}) in Eq. (\ref{simetria}) and assuming the condition $M=\partial/\partial X$, one finds
\begin{eqnarray}\nonumber
h'(x)b(x)=1, \\  \label{expresion2}
b(x)n'(x)+c(x)n(x)=0.
\end{eqnarray}
By inserting  Eq. (\ref{relacionesb}) into (\ref{expresion2}), and integrating, we obtain
\begin{eqnarray}\nonumber \label{funcionesa}
h(x) & = & \int_{0}^{x} \frac{1}{b(s)}ds,\nonumber\\ \label{funcionesb}
 n(x) & = & b(x)^{-1/2}e^{-C\int_{0}^{x}1/b(s)ds}.\nonumber
\end{eqnarray}
We can now write Eq. (\ref{estacionario}) in terms of the canonical coordinates $U$ and $X$,
\begin{equation}
\label{INLSEdis}
-\frac{d^{2}U}{dX^{2}}-2C\frac{dU}{dX}+g_{0}U^{3}-EU=0,
\end{equation}
with
\begin{equation}
\label{energia}
 E= \left(\lambda-V(x)\right) b(x)^{2}-\tfrac{1}{4}b'(x)^{2}+ \tfrac{1}{2}b(x)b''(x)+C^{2}.
 \end{equation}
Equation (\ref{INLSEdis}) is the so-called Duffing equation which arises as a model of damped nonlinear oscillations \cite{Feng}.
It follows from Eqs. (\ref{relacionesb})-(\ref{relacionesc}) that the quantity $E$ given by Eq. (\ref{energia}) is a constant of motion.

When $C=0$ the previous transformations preserve the Hamiltonian structure,
because the canonical transformation is symplectic. 
In that case, Eq. (\ref{INLSEdis}) becomes
 \begin{equation}
\label{homogenea}
-\frac{d^{2}U}{dX^{2}}+g_{0}U^3=EU.
\end{equation}
As $E$ is constant, this means that in the \emph{new variables we
obtain the nonlinear Schr\"odinger equation (NLSE) without
external potential and with an homogeneous nonlinearity}.

Of course not all choices of $V(x)$ and $g(x)$ lead to the existence
of a Lie symmetry or an appropriate canonical transformation, since they
are linked by Eqs. (\ref{relaciones}). This fact imposes some
obvious restrictions, for instance $b(x)$ must be smooth and
positive.

Note that Eq. (\ref{homogenea}) is a stationary homogeneous NLSE without external potential, which
can be reduced to quadratures and
for which many solutions are known. So, we obtain
\begin{equation}
X-X_{0}=\int_{U_{0}}^{U} \frac{dG}{\sqrt{2(N+\frac{1}{2}EG^2+\frac{1}{4}g_{0}G^4)}}, \nonumber
\end{equation}
with $N$ a constant of integration. Moreover, the energy of the system is given by
\begin{equation}
\label{energia2}
H=\frac{1}{2}\left(\frac{dU}{dX}\right)^{2}+\frac{1}{2}EU^{2}-\frac{1}{4}g_{0}U^{4}.
\end{equation}

Many solutions of Eq. (\ref{homogenea}) are known. In this
paper we will use the following ones
\begin{subequations}\label{solsol}
\begin{eqnarray}
U_1(X) & =  & \eta \frac{1} {\cosh(\mu X)}, \ \ \  \hfill \left( E=-\mu^2, g_{0}=-\frac{2\mu^2}{\eta^2}\right),  \label{sol1} \\
U_2(X)&=& \eta \tanh( \mu X), \ \
\  \left(E=2\mu^{2}, g_{0}=\frac{2\mu^{2}}{\eta^{2}}\right),\label{sol2}\\
U_3(X) & =  &\eta\frac{\sn(\mu X,k)}{\dn(\mu X,k)}, \ \ \
\left(E=\mu^{2}(1-2k^2), g_{0}=-\frac{2\mu^{2}k^{2}(1-k^{2})}{\eta^{2}}\right), \label{sol3} \\
U_4(X) & = &\eta\dn(\mu X,k), \ \ \ \left(E=\mu^{2}(k^2-2), g_{0}=-\frac{2\mu^{2}}{\eta^{2}}\right), \label{sol4}
\end{eqnarray}
\end{subequations}
with $0\leq k\leq1$.
Table \ref{tabla} summarizes the parameter values required for the existence of the solutions listed in Eqs. (\ref{solsol}).
\begin{table}[h]
\begin{tabular}{|c|c|c|c|} \hline
$U$& $E$&$g_{0}$&$H$ \\ \hline
$U_{1}(X)$ &   negative & negative &0 \\ \hline
$U_{2}(X)$ &   positive & positive    & positive  \\ \hline
$U_{3}(X)$ &   both&negative&   positive   \\ \hline
$U_{4}(X)$ &   negative&negative     &   negative  \\ \hline
\end{tabular}
\caption{ Conditions on the parameters $E$, $g_{0}$ and  energy $H$ for the existence of the solutions $U_{i},  i=1..4$ of Eq. (\ref{homogenea}) listed in Eqs. (\ref{solsol}).
\label{tabla} }
\end{table}

\section{Connection between the NLSE and INLSE via the LSE}\label{s4}

 Setting $C=0$ and eliminating $c(x)$ in Eqs.  \eqref{relaciones} we get
  \begin{equation}\label{X1}
g(x)= g_{0}/b(x)^{3},
 \end{equation}
plus an equation relating $b(x)$ and $V(x)$
 \begin{equation}\label{X2}
 b'''(x)-2b(x)V'(x)+4b'(x) \lambda -4b'(x)V(x) =0.
 \end{equation}
We notice that the simplest  form to generate solutions for our problem, which involves constructing solutions pairs $(b(x),V(x)) $ of Eq. (\ref{X2}) is to fix $b(x)$ and then, to calculate $V(x)$, since then we must solve a linear first order equation. Although we can eliminate $b(x)$ and obtain a nonlinear equation for the pairs $g(x)$ and $V(x)$ for which there is a Lie symmetry,  it is more convenient  to work with \eqref{X2}, which is a linear equation.
Alternatively, we can define $\rho(x) = b^{1/2}(x)$ and get an Ermakov-Pinney equation \cite{Ermakov,Pinney}
 \begin{equation}\label{Ek}\nonumber
 \rho_{xx} + \left( \lambda - V(x)\right)\rho = E/\rho^3,
 \end{equation}
whose solutions can be constructed as
 \begin{equation}\label{ep}\nonumber
 \rho = \left(\alpha \varphi_1^2 + 2\beta \varphi_1\varphi_2 + \gamma \varphi_2^2\right)^{1/2},
 \end{equation}
 with $\alpha, \beta, \gamma$ constant and $\varphi_j(x)$ being two linearly independent solutions of the Schr\"odinger equation
 \begin{equation}\label{final}\nonumber
 \varphi_{xx} +\left(\lambda-V(x)\right) \varphi = 0.
 \end{equation}
This choice leads to $E = \Delta W^2$ with $\Delta = \alpha \gamma - \beta^2$ and $W$ being the (constant) Wronskian $W = \varphi'_1\varphi_2 - \varphi_1 \varphi'_2$. Thus, given any arbitrary solution of the \emph{linear Schr\"odinger equation  \eqref{final} we can construct solutions of the nonlinear spatially inhomogeneous problem Eq. (\ref{estacionario}) from the known solutions of Eq. (\ref{homogenea})}. Thus, using the huge amount of knowledge on the linear
Schr\"odinger equation we can get potentials $V(x)$ for which $\varphi_1$ and $\varphi_2$ are known and
construct $b(x)$, the canonical transformations $h(x),n(x)$, the nonlinearity $g(x)$ and the explicit solutions $u(x)$.

\section{Qualitative Analysis and Exact Solutions}\label{s5}

\begin{figure}
\epsfig{file=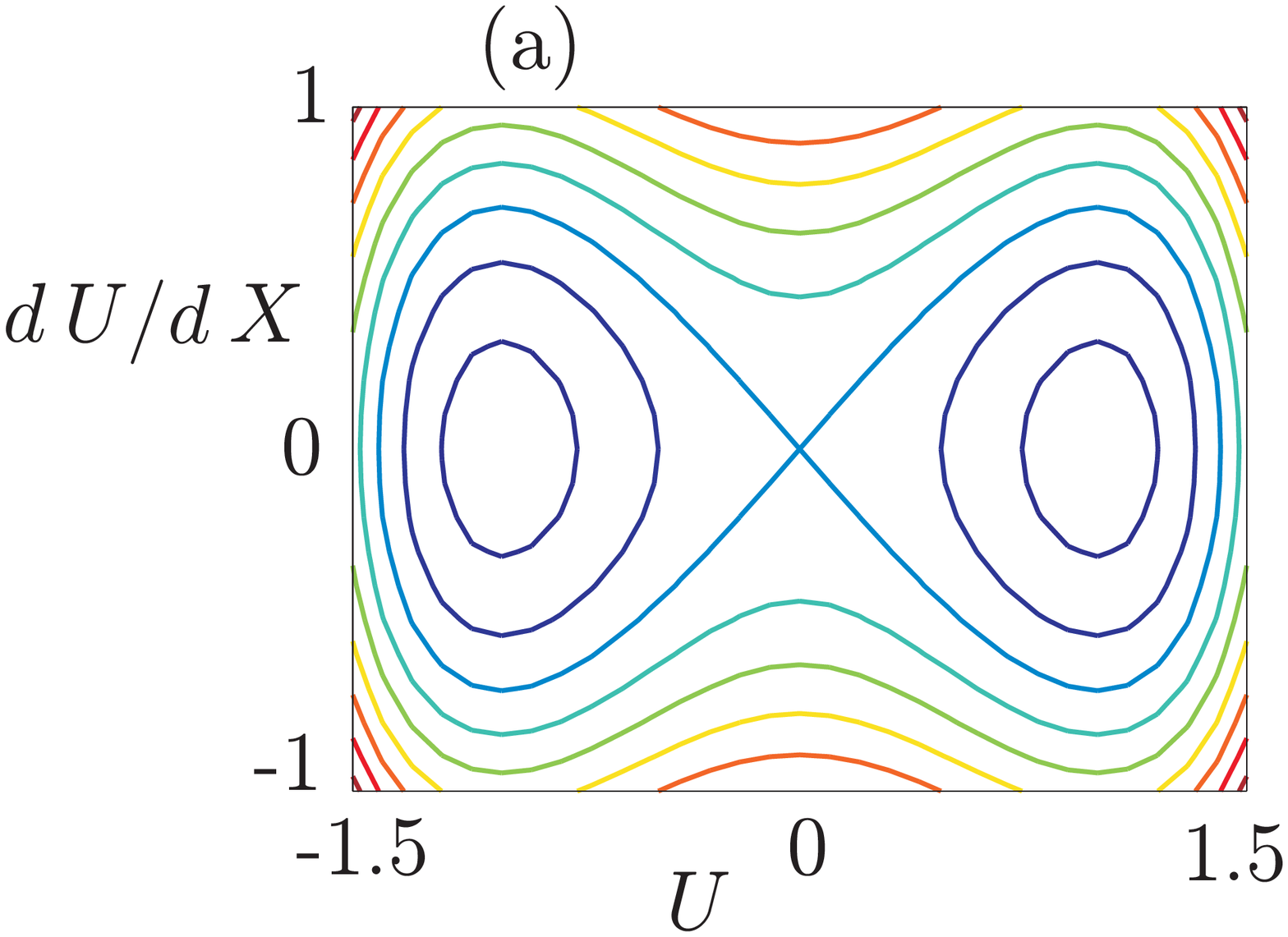,width=7cm}
\epsfig{file=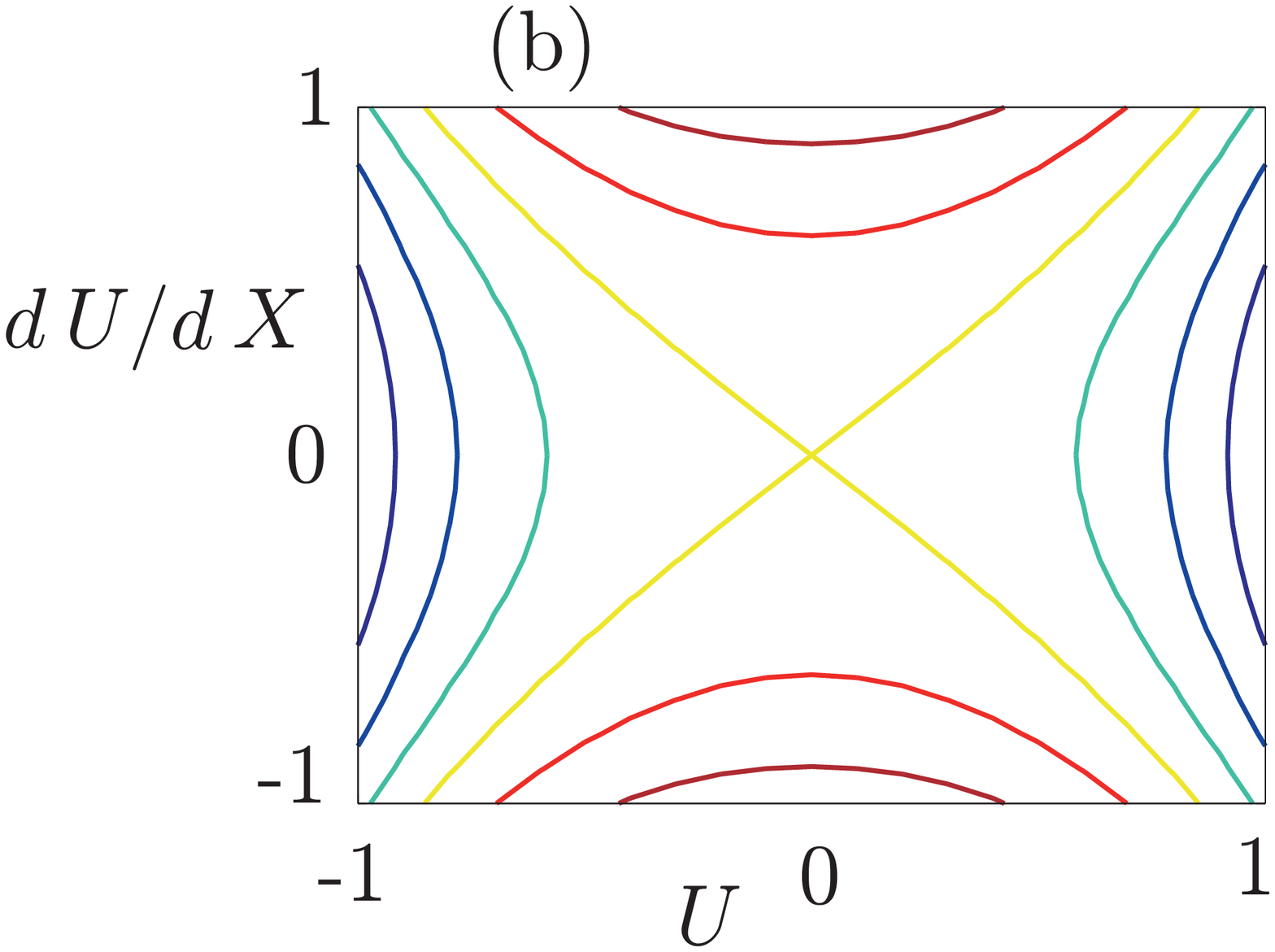,width=7cm}
\epsfig{file=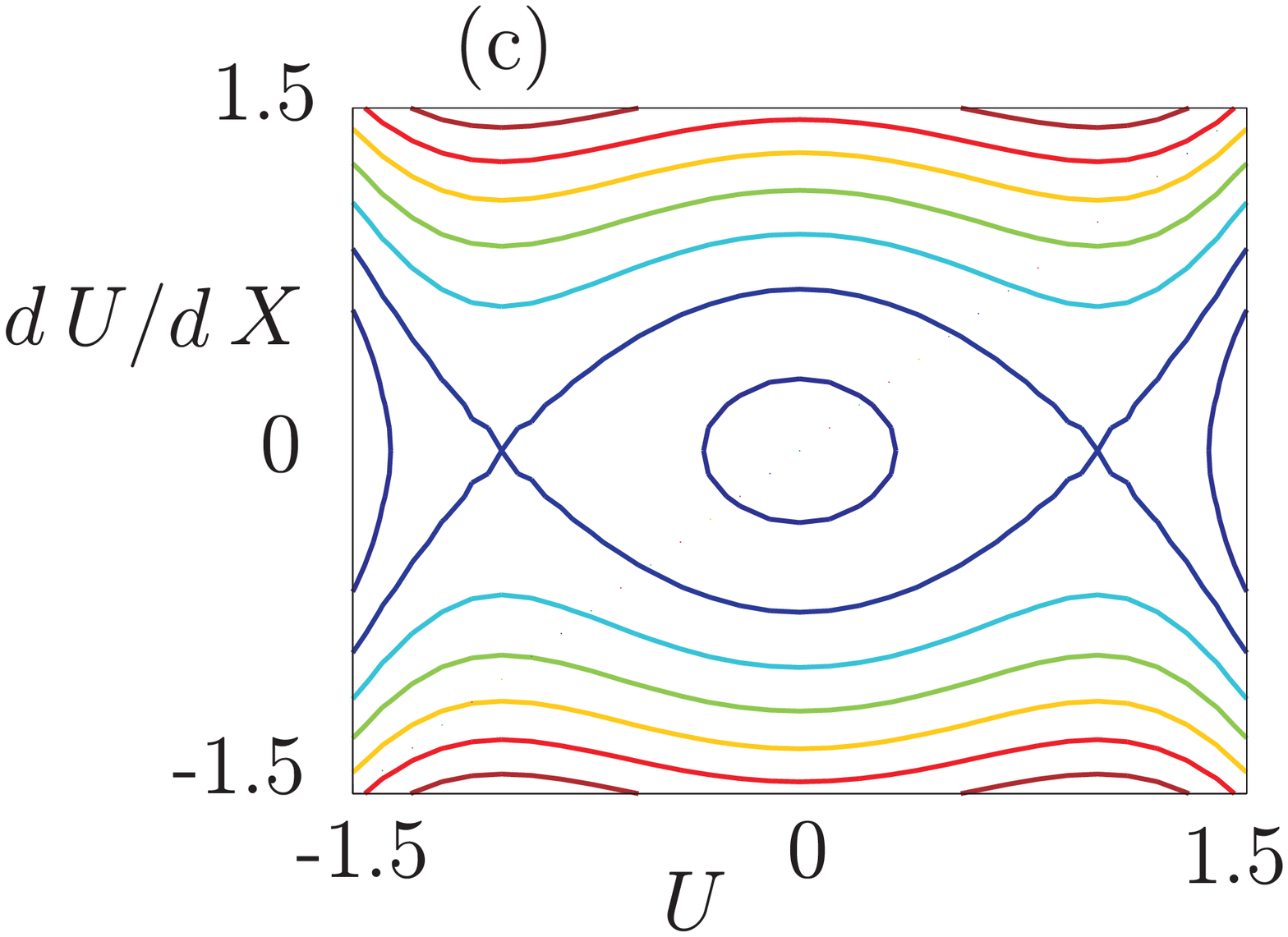,width=7cm}
\epsfig{file=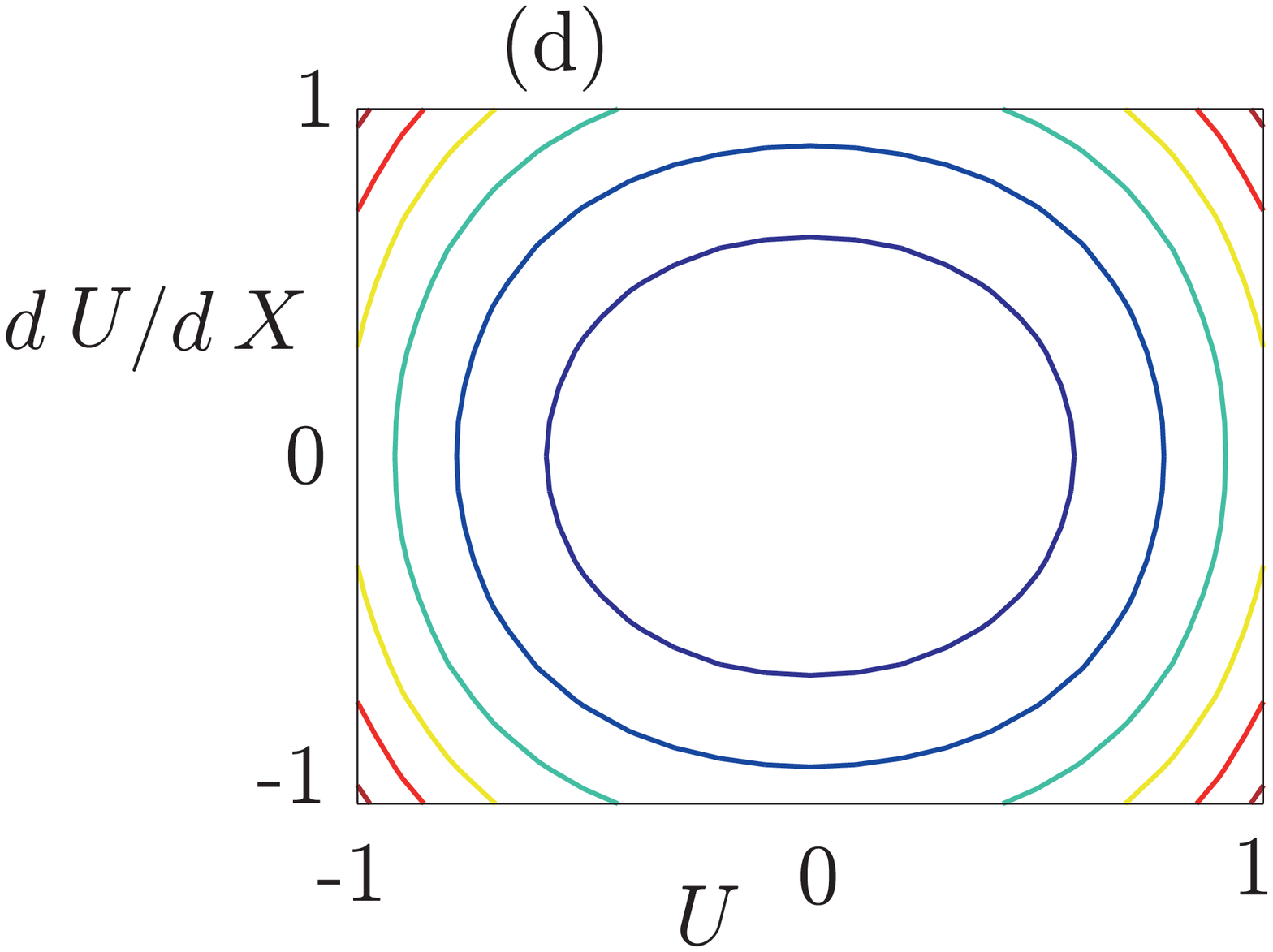,width=7cm}
\caption{[Color online]
Phase portrait of the real solutions of Eq. (\ref{homogenea}) for (a) $E<0$, $g_{0}<0$, (b) $E<0$, $g_{0}>0$ (c) $E>0$, $g_{0}>0$ and (d) $E>0$, $g_{0}<0$
\label{phase}}
\end{figure}

In this section, we will calculate exact solutions of Eq. (\ref{estacionario}) for different specific choices of the nonlinear coefficient $g(x)$ and the external potential $V(x)$, using the method described in the previous sections, for $C=0$. Moreover, using qualitative analysis, we will describe properties of the solutions of Eq. (\ref{estacionario}), on the basis of the qualitative behaviour of Eq. (\ref{homogenea}). We begin by calculating the equilibrium
points of this equation. One easily finds that the Eq.
(\ref{homogenea}) has three possible equilibrium points, depending
of the signs of $E$ and $g_{0}$:
\begin{subequations}
\begin{eqnarray}
U_{\pm} & = & \pm\sqrt{E/g_{0}},\nonumber\\ 
U & = &0. \nonumber
\end{eqnarray}
\end{subequations}%
Then, we distinguish four cases:
\begin{enumerate}
\item For $E<0, g_{0}<0$, we get  three equilibrium points. $U=0$ is a saddle point and $U_{\pm}$ are centers, Fig. \ref{phase}(a).
\item When $E<0, g_{0}>0$, we obtain that $U=0$ is the only equilibrium point, which is a saddle point, Fig. \ref{phase}(b).
\item When $E>0, g_{0}>0 $, we get three equilibrium points. $U_{\pm}$ are saddle points and $U=0$ is a center, Fig. \ref{phase}(c).
\item The last case corresponds to $E>0, g_{0}<0$. For this case, the only equilibrium point is the trivial solution $U=0$, which is a global center, Fig. \ref{phase}(d).
\end{enumerate}
Using Eq. (\ref{energia2}), we can draw the phase portrait of Eq.
(\ref{homogenea}), as we can see in Fig. \ref{phase}.
 \begin{figure}
\epsfig{file=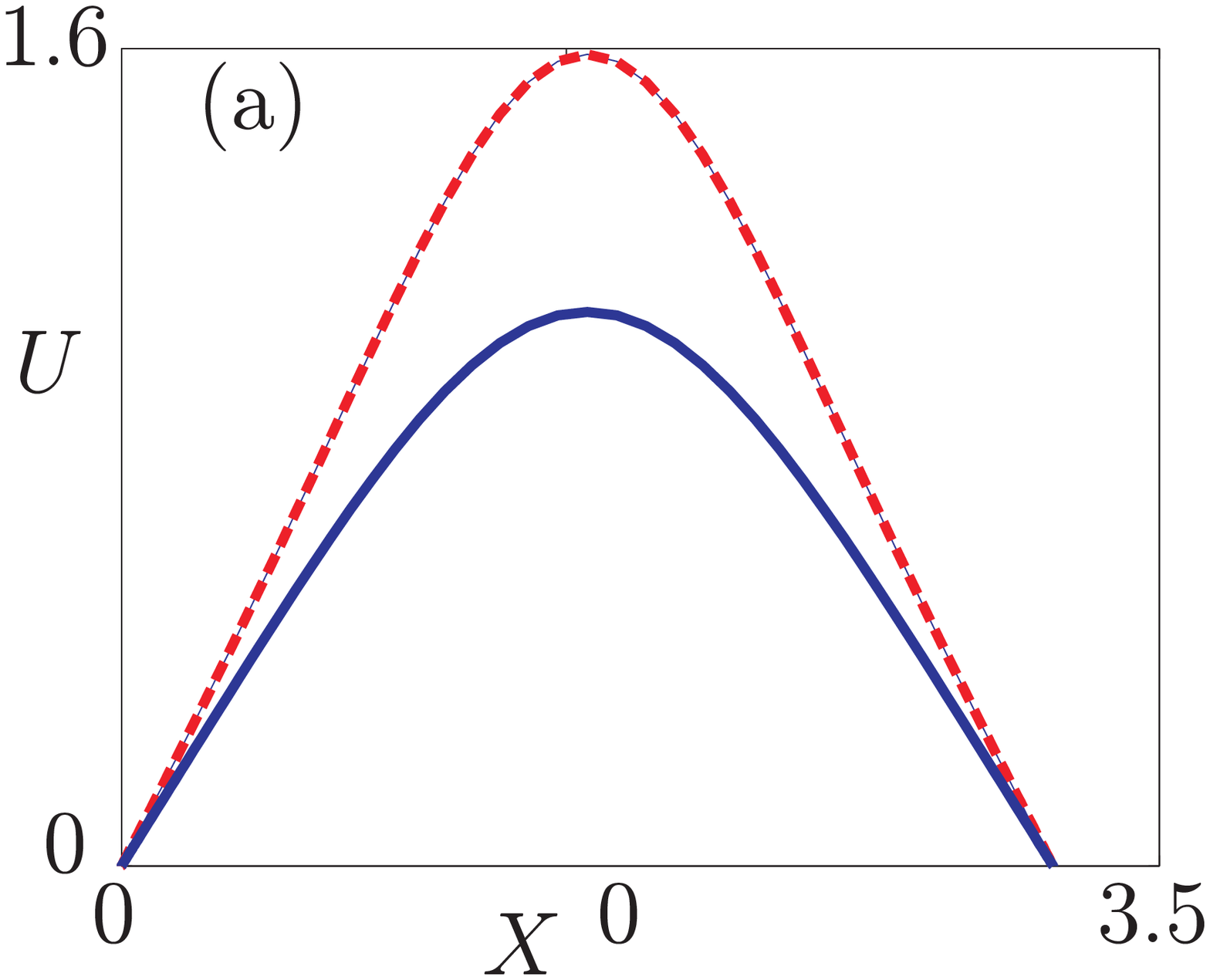,width=7cm}
\epsfig{file=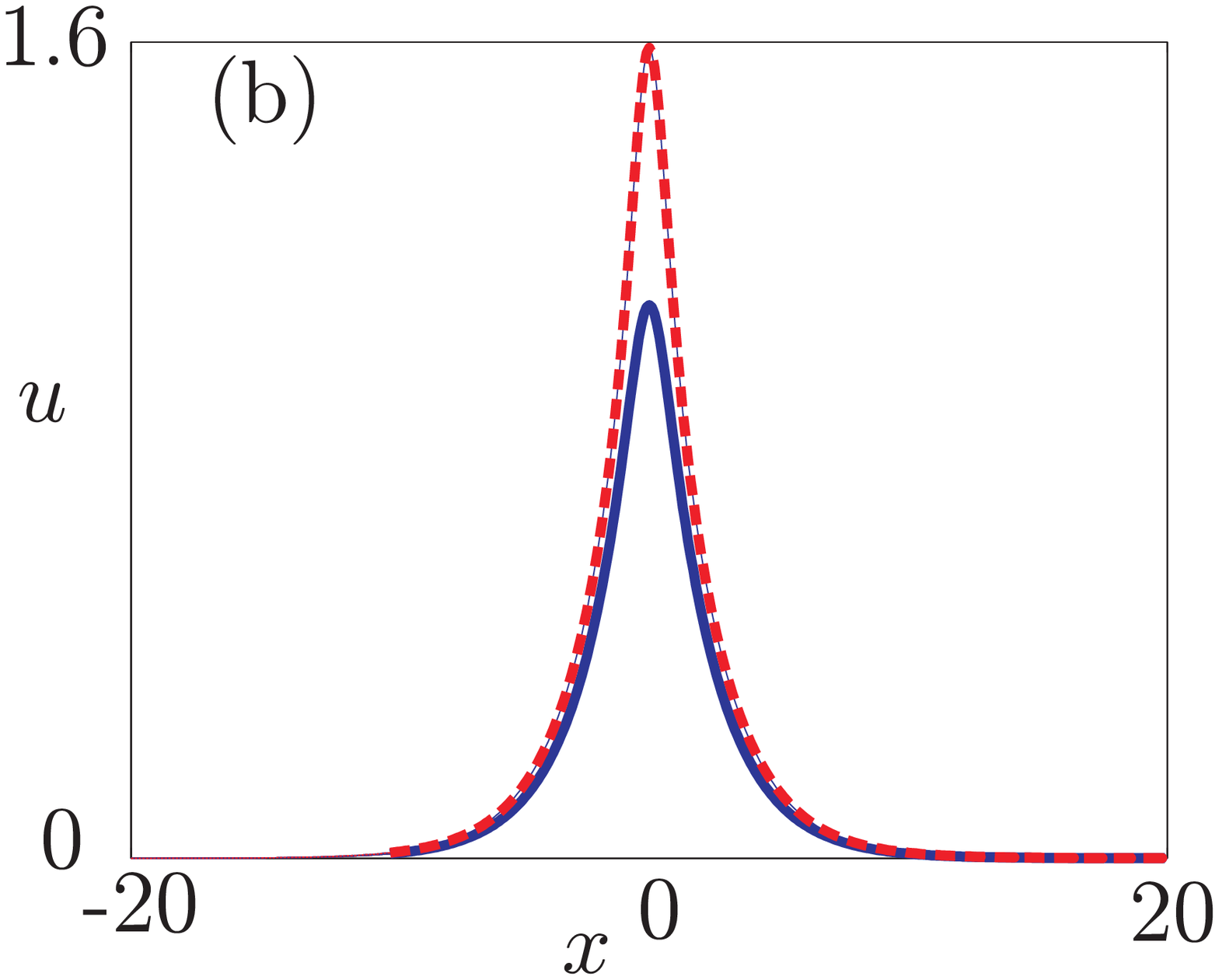,width=7cm}
\caption{[Color online]
 Solutions of (a) Eq. (\ref{homogenea}) and (b) Eq. (\ref{estacionario}), for $E=0.15$, $g_{0}=-1$ (solid blue line) and $E=-0.75$, $g_{0}=-1$ (dashed red line) in both cases. We apply the transformation  (\ref{transformacion}) to the solutions of Eq.  (\ref{homogenea}) shown in Fig. \ref{bright}(a) to obtain  the solutions of Eq. (\ref{estacionario}) shown in Fig \ref{bright}(b).
\label{bright}}
\end{figure}

In what follows we will present  three examples as applications of our theory:
\bigskip

{\noindent \bf Example 1.} Let us take $b(x)=\cosh(x)$. By using
Eqs. (\ref{X2}) and (\ref{energia}), for $C=0$, we obtain
\begin{equation}
V(x)=\lambda+\frac{1}{4}+\left(\frac{1}{4}-E\right)\frac{1}{\cosh^{2}(x)}. \nonumber
\end{equation}
Moreover, using Eq. (\ref{X1}), $g(x)$ is given by
\begin{equation}
g(x)=\frac{g_{0}}{\cosh^{3}(x)}, \nonumber
\end{equation}
with $X(x)$ being
\begin{equation}
\cos X(x)=-\tanh x, \nonumber
\end{equation}
where $0\leq X\leq\pi$, subject to the Dirichlet boundary conditions $U(0)=U(\pi)=0$. Any solution $U$ of Eq.
(\ref{homogenea}) gives a solution
\begin{equation}\nonumber
u(x)=b^{1/2}(x)U(X(x)),
\end{equation}
of the original equation (\ref{estacionario}). If $E<0$ and $g_{0}<0$, we are in the first case, Fig. \ref{phase}(a). The periodic solution ($\ref{sol3}$) of Eq.
(\ref{homogenea}) is a closed orbit of the phase portrait
$(U,dU/dX)$, corresponding to one of the external closed orbits to the homoclinic orbits, shown in Fig. \ref{phase}(a). By an
elementary application of L'Hopital rule, it is easy to verify
that the solution
\begin{equation}\label{transformacion} 
u(x)=b^{1/2}(x)U_{3}(X(x)),
\end{equation}
is a homoclinic orbit to zero (bright soliton) of the original equation (\ref{estacionario}). 

If $E>0$ and $g_{0}<0$, we are in the four case,
where $U=0$ is a center, Fig. \ref{phase}(d). Again,
$u(x)=b^{1/2}(x)U_{3}(X(x))$ is a homoclinic orbit (bright
soliton). The solutions of Eq. (\ref{estacionario}) for $E=0.15$, $g_{0}=-1$ and $E=-0.75$, $g_{0}=-1$ are drawn in Fig. \ref{bright} (b). The case $E=1/4$, where $V(x)$ is a constant, was studied in \cite{Nuestro}.

\bigskip
{\noindent \bf Example 2.} Let us take $V(x)=0$. Then Eq. (\ref{X2})
becomes $b'''(x)+4\lambda b'(x)=0$. For $\lambda>0$, the solution
can be written as
\begin{equation}
b(x)=1+\alpha \cos(2\sqrt{\lambda}x). \nonumber
\end{equation}
Using Eq. (\ref{X1}), we obtain a periodic nonlinearity
\begin{equation}\label{qper}
g(x)=g_{0}(1+\alpha\cos(2\sqrt{\lambda}x))^{-3}. \nonumber
\end{equation}
For small $\alpha$, this nonlinearity is approximately harmonic
\begin{equation}
g(x)\simeq g_{0}(1-3\alpha\cos(2\sqrt{\lambda}x)),\ \ \ \alpha\ll 1. \nonumber
\end{equation}
We can construct our canonical transformation by using
Eqs. (\ref{transformaciones}) and obtain
 \begin{equation}\label{cuci}
 X(x)  = \frac{1}{\sqrt{\lambda(1-\alpha^2)}}\arctan \left(
  \sqrt{ \frac{1-\alpha}{1+\alpha} } \; \tan (\sqrt{\lambda} x)\right).
\end{equation}
\begin{figure}
\epsfig{file=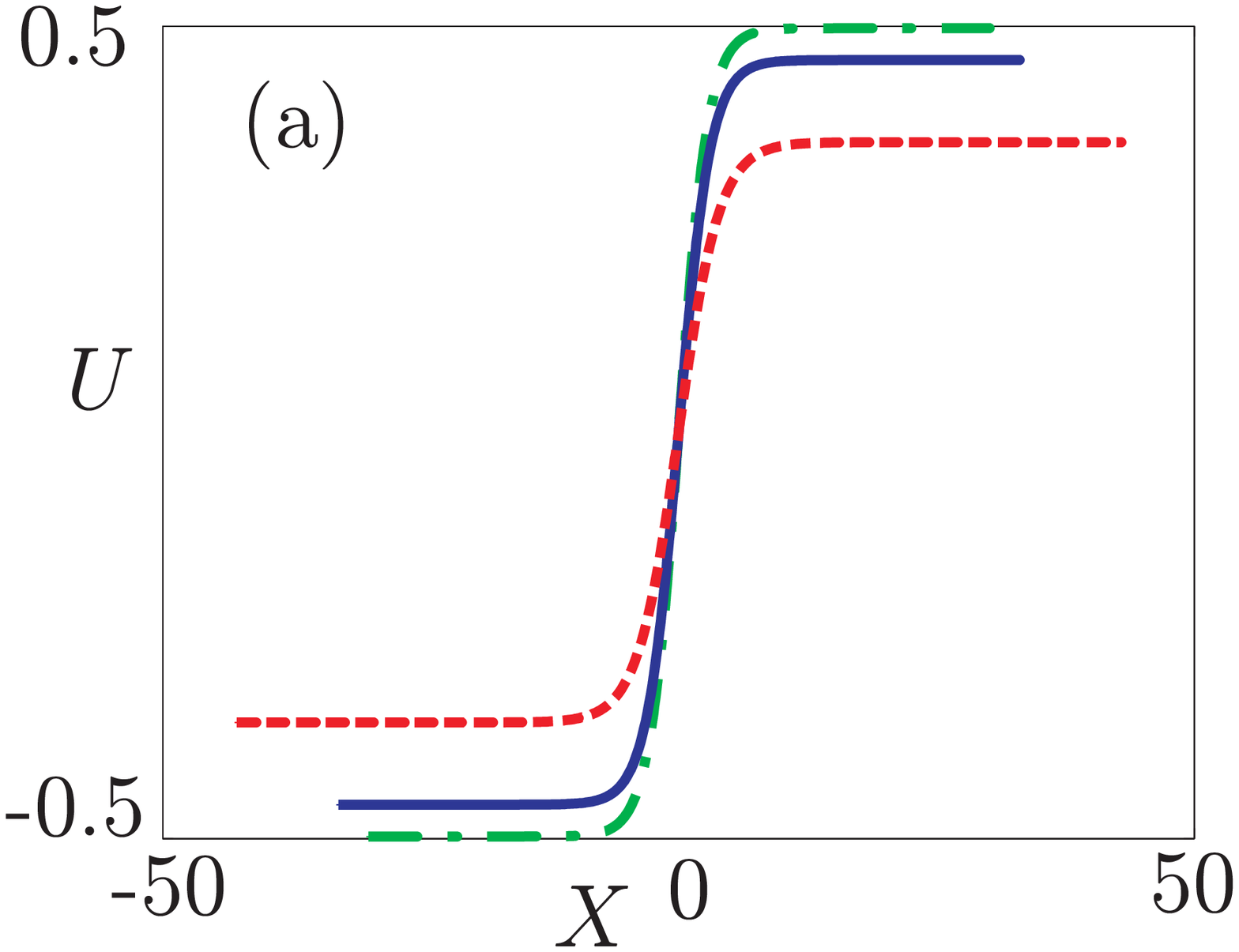,width=7cm}
\epsfig{file=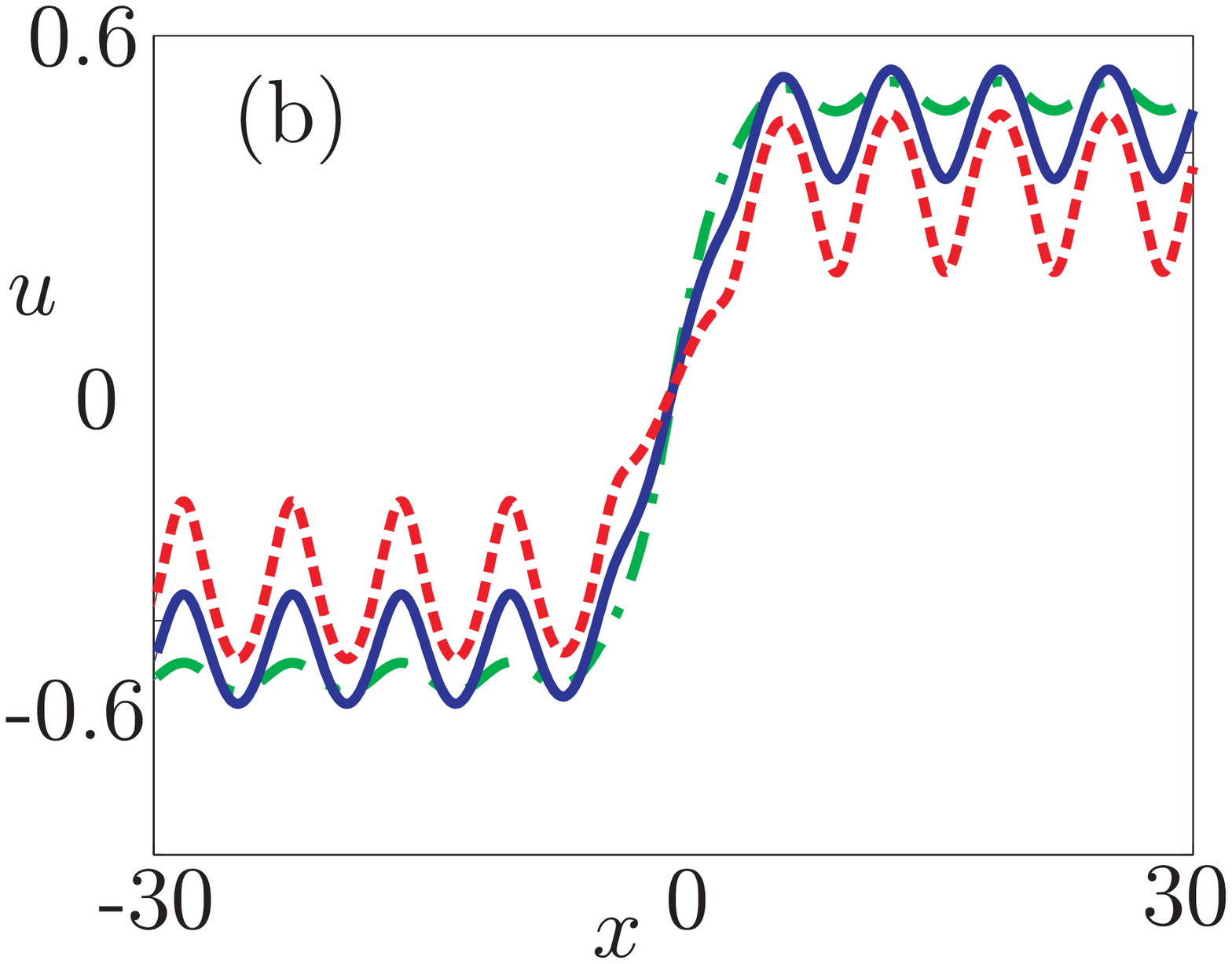,width=7cm}
\caption{[Color online] Example black solitons solutions of (a) Eq. (\ref{homogenea}) and (b) Eq. (\ref{estacionario}) with $g_{0}=1$, $\lambda=1/4$ and (i) $\alpha=0.1$ (dashed-dot green line), (ii) $\alpha=0.4$ (solid blue line) and (iii) $\alpha=0.7$ (dashed red line).The solutions shown in Fig. \ref{dark}(b) are obtained from those shown in Fig. \ref{dark}(a) through the transformations (\ref{transformaciones}).
\label{dark}}
\end{figure}
Using any solution of Eq. (\ref{homogenea}) with
$E=\lambda\left(1-\alpha^2\right)$ this transformation provides
solutions of Eq. (\ref{estacionario}) with $g(x)$ given by
\eqref{qper}. For example, when $g_{0}>0$ we can use $U_2$ as
defined by (\ref{sol2}), which in the phase portrait is the
heteroclinic orbit shown in Fig. \ref{phase}(c). So, the solution
of Eq. (\ref{estacionario}) is
 of the form
 \begin{equation}\label{brito}
 u(x)  =
 \sqrt{ \frac{\lambda(1-\alpha^2)}{g_0}
 \left(1 + \alpha \cos (2 \sqrt{\lambda} x) \right)}
\times  \tanh\left[
    \sqrt{ \frac{\lambda(1-\alpha^2)}{2} } \; X(x)
 \right].
 \end{equation}
As we can see in Fig. \ref{dark}(b), $u(x)$ is a heteroclinic
connection between periodic solutions of Eq. (\ref{estacionario}).
It is important to note that in the asymptotic regions the profile
of $u(x)$ is close to $b^{1/2}(x)$ multiplied by a constant.
Therefore, the canonical transformation (\ref{transformaciones}),
in this case, transforms a heteroclinic orbit in the phase
portrait $(U,dU/dX)$ into a heteroclinic connection, Eq.
(\ref{brito}). On the other hand, any closed orbit $U$ inside the
heteroclinic loop of the phase portrait $(U,dU/dX)$ provides a new
heteroclinic connection of the original equation
(\ref{estacionario}).

\bigskip
{\noindent \bf Example 3.} The last example is the so-called
quasi-harmonic confinement $V(x)\sim x^2$.

\begin{figure}
\epsfig{file=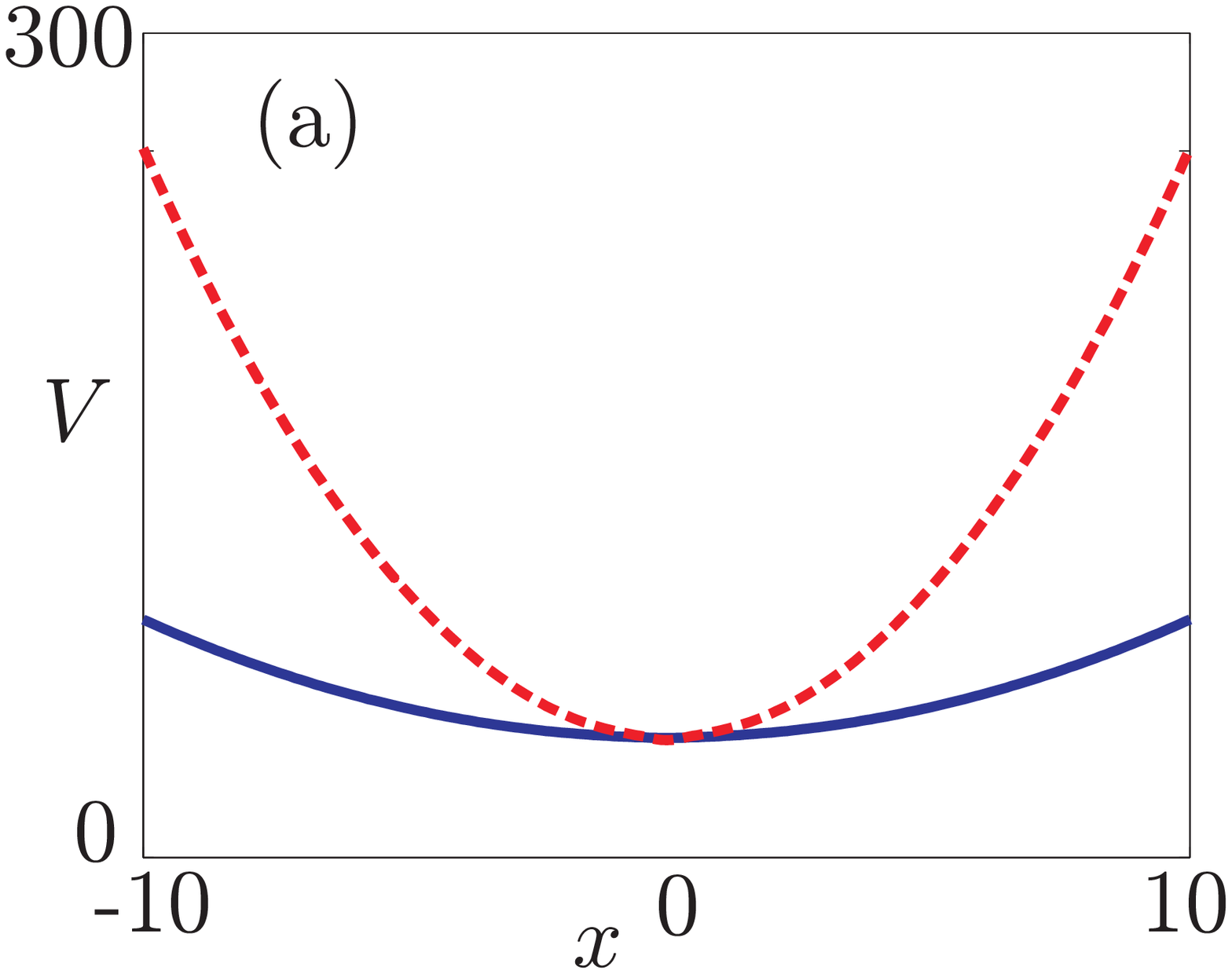,width=7cm}
\epsfig{file=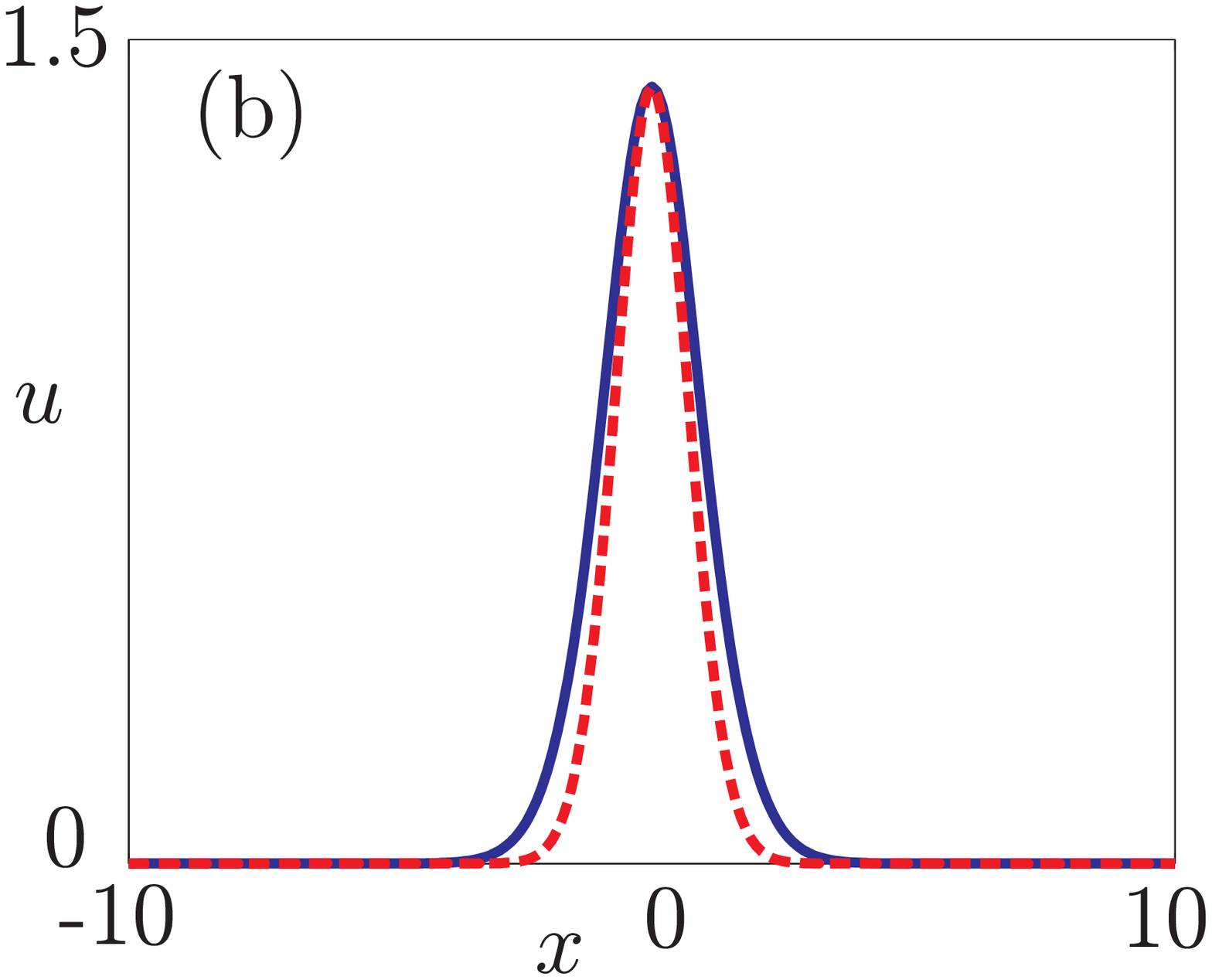,width=7cm}
\caption{[Color online]
(a) Quasi-harmonic potential for $M=1$, $\lambda=1$ and (i) $\beta=0.5$ (solid blue line) and (ii) $\beta=2.5$ (dashed red line). (b) Solutions of Eq. (\ref{estacionario}) for $\alpha=1$, $g_{0}=-1$, $M=1$ and (i) $\beta=0.5$ (solid blue line) and (ii) $\beta=2.5$ (dashed red line)
\label{solitonbright}}
\end{figure}

 If we choose $b(x)=\alpha/\sqrt{1+\beta x^2}$, with $\alpha, \beta>0 $, then, we obtain
 \begin{equation}
 g(x)=\frac{g_{0}}{\alpha^3}(1+\beta x^2)^{3/2},\nonumber
 \end{equation}
 and
 \begin{equation}
 V(x)=M(1+\beta x^2)+\frac{1}{4}\frac{3\beta x^2-2\beta+4\lambda+8\lambda\beta x^2+4\lambda\beta^{2}x^{4}}{(1+\beta x^2)^{2}}, \nonumber
 \end{equation}
 with $M$ a positive constant. Although the expression of $V(x)$ is complicated this potential $V(x)$ is a quasi-harmonic potential and satisfies $V(x) \sim x^2$ for large $x$. Moreover $V(x)$ is a harmonic potential with a bounded perturbative term (see Fig. \ref{solitonbright}(a)). As to the nonlinear term, it satisfies, $g(x)\sim x^{2}$ for $|x|\ll 1$,  and $g(x)\sim x^{3}$ for $|x| \gg 1$.
 Using Eq. (\ref{energia})
we get $E=-\alpha^2M$. Taking $g_{0}<0$, we obtain the nonlinear
Schr\"odinger equation with nonlinear attractive term, Eq. (\ref{homogenea}).
As $E<0$ and $g_{0}<0$,
all the solutions
of Eq. (\ref{homogenea}) are bounded, (see Fig. \ref{phase}(a)). If $U(X)$ is one of these solutions,
 it is clear that
\begin{equation}
 u(x)=b(x)^{1/2}U(X(x)), \nonumber
 \end{equation}
 is a homoclinic orbit (bright soliton) of the original
equation. In particular, the solution given by Eq. (\ref{sol1}) is
 \begin{equation}
 U_{1}(X)=\sqrt{\frac{2E}{g_{0}}} \frac{1}{\cosh(\sqrt{|E|}X)}. \nonumber
 \end{equation}
 As $X(x)=x\sqrt{1+\beta x^{2}}/(2\alpha)+\sinh^{-1}(\sqrt{\beta}x)/(2\alpha\sqrt{\beta})  $, we get
 \begin{equation}
 u(x)=b(x)^{1/2}U_{1}(X(x)). \nonumber
 \end{equation}
In Fig. \ref{solitonbright}(b), we draw the solutions of Eq. (\ref{estacionario}) for different values of the parameter $\beta$.

In this example, we have used the solution $U_{1}(X)$ of Eq. (\ref{homogenea}). Another possibility
is to choose the closed periodic orbits inside the homoclinic loop (see Fig. \ref{phase}(a))
given by $U_{4}(X)$ in
Eq. (\ref{sol4}). We note that, for this case, Eq. (\ref{energia2}) satisfies $H\leq 0$, as one can see in Table \ref{tabla}.

The analytical expression of the closed periodic orbits outside the homoclinic loop is given by $U_{3}(X)$, with $k>1/\sqrt{2}$. Such orbits satisfy $H>0$ (see Table \ref{tabla}).

\begin{figure}
\epsfig{file=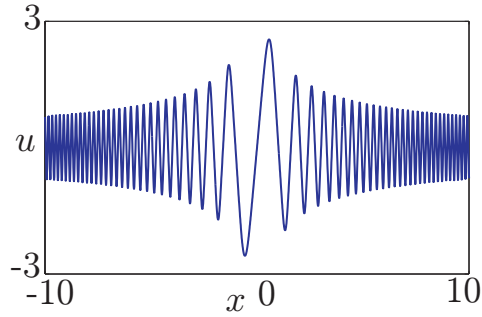,width=7cm}
\caption{Solution of Eq. (\ref{estacionario})  with infinite nodes for $\alpha= 1$, $\beta=2.5$, $g_{0}=-1$ and $k=3/4$
\label{infinito}}
\end{figure}

As $X$ is a bijective map on the real line, $b(x)$ is a positive function and $U_{3}(X)$ is a periodic function with infinite nodes on the real line, the function $u(x)=b(x)^{1/2}U_{3}(X)$ also has infinite nodes on the real line, as it is shown in Fig. \ref{infinito}.
It is immediate to check that the solution $u\rightarrow 0$ when $x\rightarrow \pm\infty$. Moreover, the zeroes of $\phi$ accumulate for large values of $x$, since the distance between two consecutive zeroes is given by $x_{n+1}-x_{n}\sim\sqrt{n+1}-\sqrt{n}$. Other localized solutions with infinite nodes have been studied in a different context in Ref. \cite{infnodos2}.

\section{Asymmetric modes of the INLSE}\label{s6}

In the previous section, we have explored the case $C=0$ in the
canonical transformation (\ref{transformaciones}). In this case, the original equation (\ref{estacionario}) is
reduced to a Nonlinear Schr\"odinger Equation. In this section, we will study the
case $C>0$. If we take $g_{0}<0$ and $E>0$, the
resulting equation is
\begin{equation}
\label{INLSEdisneg}
\frac{d^{2}U}{dX^{2}}+2C\frac{dU}{dX}+|g_{0}|U^{3}+EU=0.
\end{equation}
In general, this equation is not integrable and the energy is not
a conserved quantity. However, exact solutions of Eq. (\ref{INLSEdisneg}) can be
constructed  analytically in particular cases.  An exact integrability condition was given in
\cite{Gendelman}
\begin{equation}\label{integrability} \nonumber
 E=\frac{8}{9}C^{2}.
\end{equation}
In that case  a family of exact analytical solutions of Eq. (\ref{INLSEdisneg}) is given by the expression
\begin{equation}
\label{solucionsninicial}\nonumber
U_{n}(X)=\frac{\mu_{n}}{\sqrt{2|g_{0}|}}e^{-BX}\frac{\sn\left(\frac{\mu_{n}}{B}(1-e^{-BX}), \sqrt{2}/2\right)}{\dn\left(\frac{\mu_{n}}{B}(1-e^{-BX}), \sqrt{2}/2 \right)}, \quad
 n=1,2,3,...
\end{equation}
where $\mu_{n}$ and $B$ are related to the boundary conditions of the problem .

We are going to solve Eq. (\ref{estacionario}), using the solutions of Eq. (\ref{INLSEdisneg}).
Choosing $b(x)=\cosh(x)$ and using Eqs. (\ref{X2}) and (\ref{energia}), we can calculate the potential $V(x)$:
\begin{equation}\label{V1}\nonumber
V(x)=\lambda+1/4+\left(\frac{1}{4}+\frac{C^{2}}{9}\right)\frac{1}{\cosh^{2}(x)}.
\end{equation}
 The nonlinear term is
  \begin{equation}\label{nolineal}\nonumber
 g(x)=g_{0}\cosh^{-3}(x)e^{-2CX(x)}.
 \end{equation}
 So, we can calculate the solutions of Eq. (\ref{estacionario}) for the case $C\neq 0$ and to compare them with the solution obtained for the case $C=0$, example 1.
 In this way, we can construct our canonical transformation by using Eqs. (\ref{transformaciones}) and obtain
 \begin{equation}\nonumber
 \cos X(x)=-\tanh x.
 \end{equation}
 Then, $0 \leq X \leq\pi$ and using the boundary conditions for $u$, $\lim_{|x|\rightarrow\infty}u(x)=0$, one has to impose $U(0)=U(\pi)=0$.

 \begin{figure}
 \epsfig{file=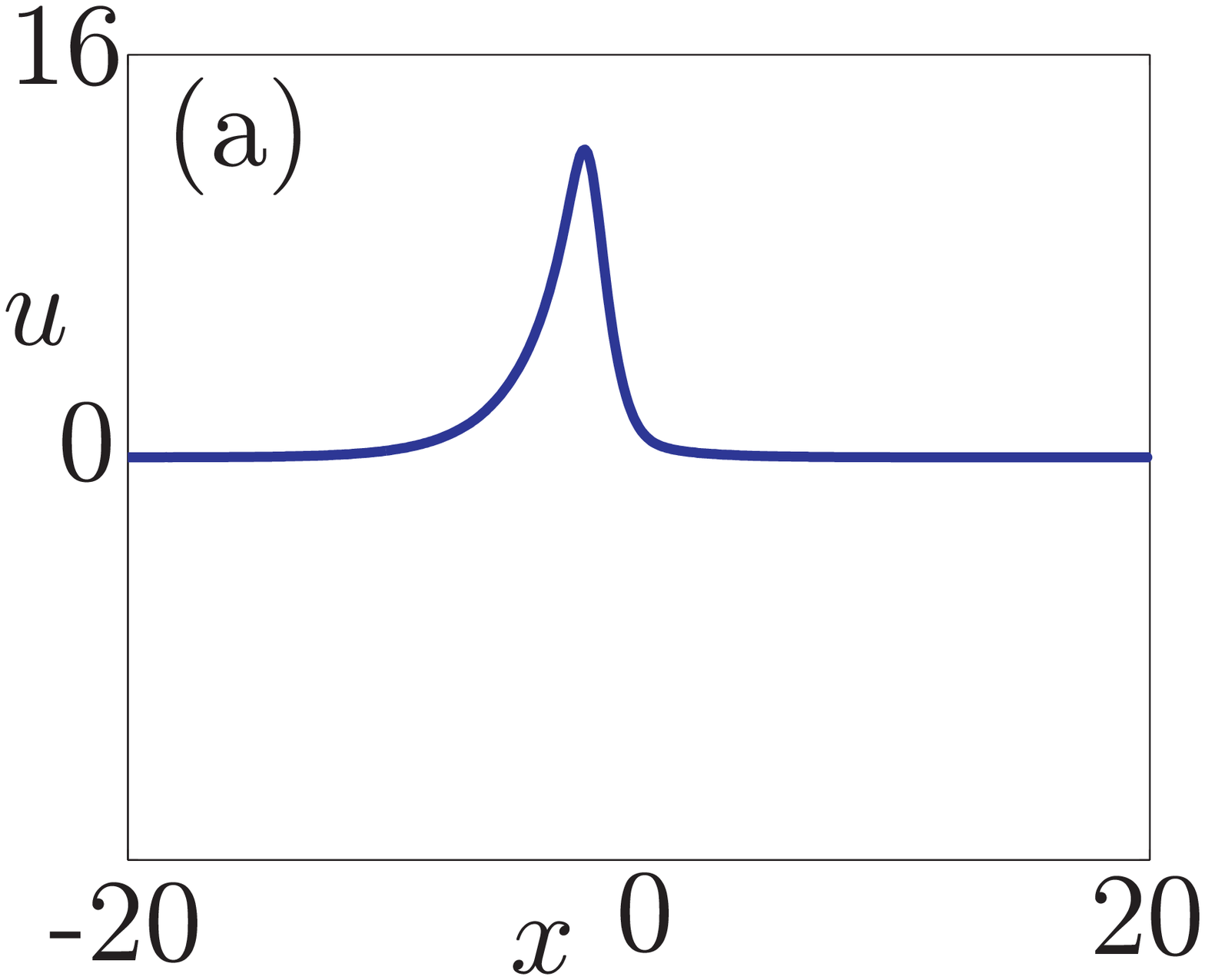,width=4.5cm}
 \epsfig{file=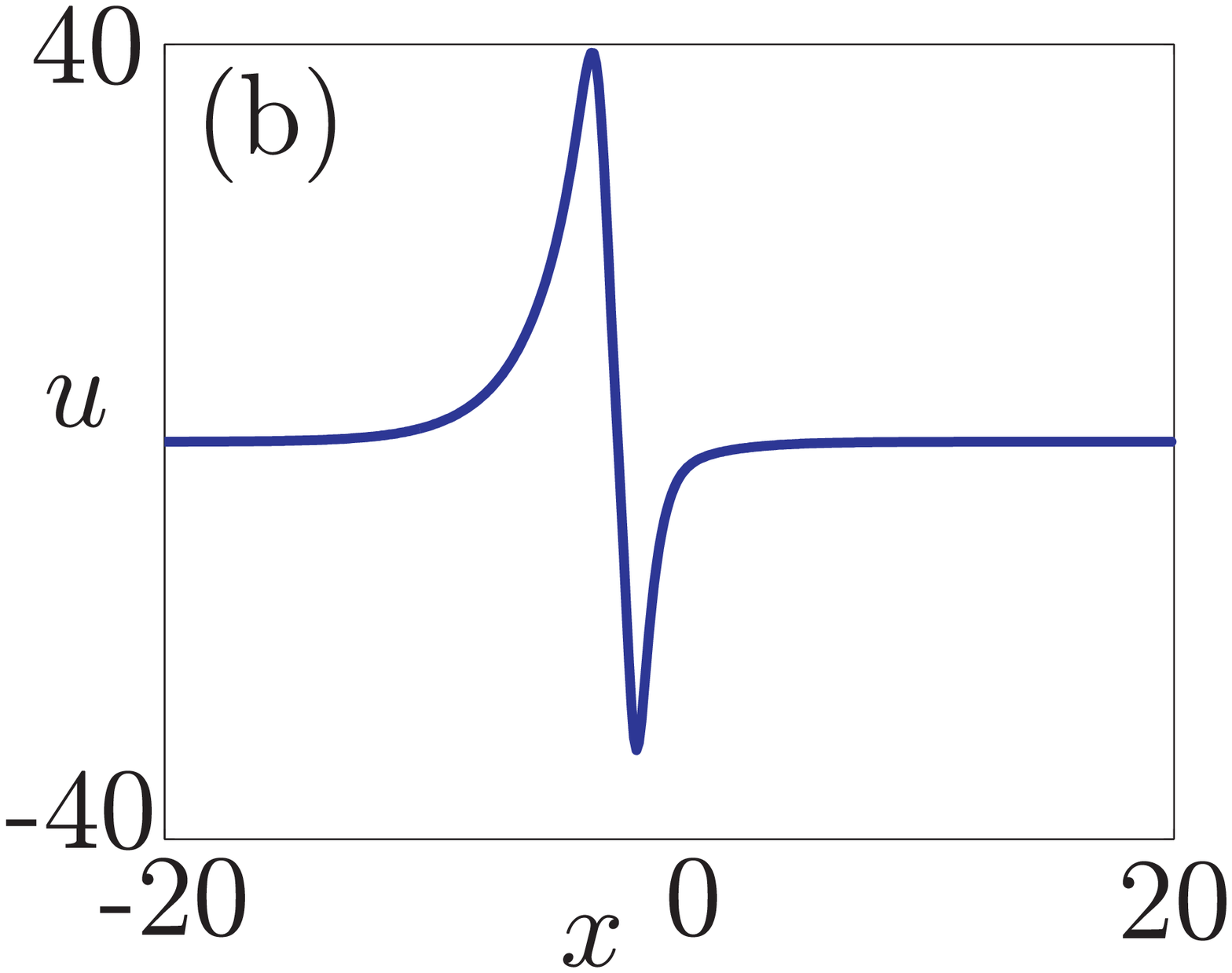,width=4.5cm}
 \epsfig{file=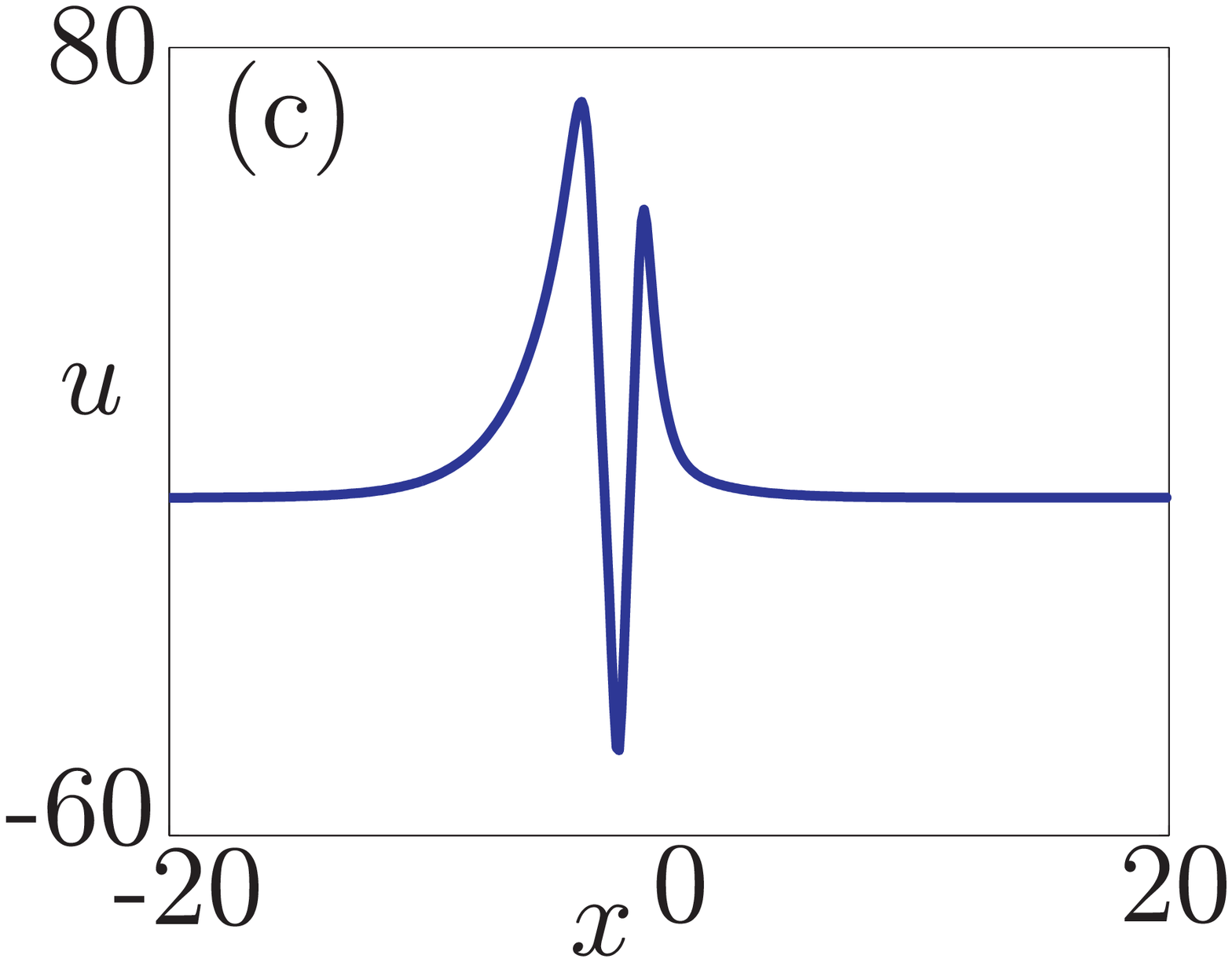,width=4.5cm}
 \caption{Asymmetric solutions of Eq. (\ref{estacionario}) with $C=4$ for (a) $n=1$, (b) $n=2$ and (c) $n=3$.
 \label{osc}}
 \end{figure}

 Using these boundary conditions, we obtain the value of the amplitude $\mu_{n}$  as a function of the value of the integer $n$
 \begin{eqnarray}\nonumber
\mu_{n}=\frac{4CK(\sqrt{2}/2)}{3(1-e^{-2C\pi/3})}n, \nonumber
\end{eqnarray}
and $B=2C/3$. Thus, the solutions of Eq. (\ref{INLSEdisneg}) are
\begin{eqnarray}
\label{solucionsn}\nonumber
U_{n}(X)=\frac{\mu_{n}}{\sqrt{2|g_{0}|}}e^{-2CX/3}\frac{\sn\left(2nK(\sqrt{2}/2)(1-e^{-2CX/3})/(1-e^{-2C\pi/3}), \sqrt{2}/2\right)}{\dn\left(2nK(\sqrt{2}/2)(1-e^{-2CX/3})/(1-e^{-2C\pi/3}), \sqrt{2}/2 \right)}, \\ \nonumber
 n=1,2,3,...
\end{eqnarray}
where $K(k)$ is the elliptic integral of the first kind,
\begin{equation}
K(k)=\int_{0}^{\pi/2}\frac{d\theta}{\sqrt{1-k^{2}\sin^{2}(\theta)}}. \nonumber
\end{equation}
Then, the solutions of Eq. (\ref{estacionario}) are
\begin{equation}
\label{sf}
u_{n}(x)=b^{1/2}(x)e^{CX(x)}U_{n}(X(x)),\quad n=1,2,... 
\end{equation}
 By using L'Hopital's rule, $u_{n}(x)\rightarrow0$ when
$|x|\rightarrow\infty$ in (\ref{sf}). So, the solutions (\ref{sf}) are
localized solutions of our problem as it can be seen in Fig.
\ref{osc}. These solutions are asymmetric solutions of Eq.
(\ref{estacionario}). Moreover, each of those solutions has exactly $n-1$ zeroes. In Fig. \ref{osc}, we plot some of them corresponding to $n=1,2,3$. The picture in Fig. \ref{osc}(a) shows clearly the difference between the positive solution given by (\ref{sf}), for $C\neq 0$, and the positive solution plotted in Fig. \ref{bright}(b) and given by Eq. (\ref{transformacion}), for $C=0$.

\section{Conclusions}\label{s7}

In this paper, we have used the method of Lie symmetries to find exact solutions of the INLSE. We have introduced the general framework of the Lie's theory and presented different examples as application to the theory. By using the qualitative theory of the dynamical systems, we can show the properties of the solutions of the INLSE and to classify such solutions. Finally, we have calculated asymetric solitons of the inhomogeneous nonlinear Schr\"odinger equation.

\section*{Acknowledgments}

This work has been partially supported by grants FIS2006-04190, MTM2005-03483
 (Ministerio de Educaci\'on y Ciencia, Spain), PAI-05-001 and PCI08-093 (Consejer\'{\i}a de Educaci\'on y Ciencia de la Junta de Comunidades de Castilla-La Mancha, Spain).

\end{document}